# Percolation-induced resistivity drop in cold-pressed LuH$_2$

Ningning Wang[1,2#*], Jun Hou[1,2#], Ziyi Liu[1,2], Pengfei Shan[1,2], Congcong Chai[1,2], Shifeng Jin[1,2], Xiao Wang[1,2], Youwen Long[1,2], Yue Liu[1,2], Hua Zhang[1,2], Xiaoli Dong[1,2], and Jinguang Cheng[1,2*]

[1]Beijing National Laboratory for Condensed Matter Physics and Institute of Physics, Chinese Academy of Sciences, Beijing 100190, China

[2]School of Physical Sciences, University of Chinese Academy of Sciences, Beijing 100190, China

\# These authors contributed equally to this work.

*Corresponding authors: nnwang@iphy.ac.cn; jgcheng@iphy.ac.cn

## Abstract

The stoichiometric bulk LuH$_2$ is a paramagnetic metal with high electrical conductivity comparable to simple metals. Here we show that the resistivity of cold-pressed (CP) LuH$_2$ samples varies sensitively upon modifying the grain size or surface conditions via the grinding process, *i.e.*, the CP pellets made of commercially purchased LuH$_2$ powder remain metallic but exhibit thousands of times higher resistivity, while additional grinding of LuH$_2$ powders in air further enhances the resistivity and even results in weakly localized behaviors. For these CP samples, interestingly, we can *occasionally* observe abrupt resistivity drops at high temperatures, which also show dependences on magnetic fields and electrical current. Measurements of variable-temperature XRD, magnetic susceptibility, and specific heat exclude the possibilities of structural, magnetic, and superconducting transitions for the observed resistivity drops. Instead, we tentatively attribute these above observations to the presence of insulating layers on the grain surface due to the modification of hydrogen stoichiometry or the pollution by oxygen/nitrogen. Percolation of the metallic grains through the insulating surfaces can explain the sudden drop in resistivity. The present results thus call for caution in asserting the resistivity drops as superconductivity and invalidate the background subtraction in analyzing the resistivity data.

## Introduction

The recent report of near-ambient superconductivity in a nitrogen (N)-doped lutetium (Lu) hydride [1] has aroused considerable interest because it, if proven to be true, would represent a milestone towards room-temperature superconductivity. According to this report, the sample recovered from the reaction of a thin Lu foil with $H_2/N_2$ (99:1) mixture at 2 GPa and 65 °C consists of a major cubic $LuH_{3-\delta}N_\varepsilon$ (92.25%) with the minor phases of $LuN_{1-\delta}H_\varepsilon$ (7.29%) and $Lu_2O_3$ (0.46%). In addition to a high superconducting critical temperature $T_c \approx 294$ K at 1 GPa, the sample was observed to exhibit visible color changes from blue through pink to bright red upon compression. The high-$T_c$ superconductivity was found to exist only in the pink-colored phase over the pressure range of 0.3-3 GPa and seems to be supported by the electrical transport, magnetic, and thermodynamic measurements in Ref. [1].

However, subsequent follow-up studies immediately cast doubts on the validity of these results[2-5]. For example, a density-functional-theory study on the structural stability and optical absorption indicated that the parent structure of the N-doped Lu hydride should be $LuH_2$ with the cubic fluorite structure rather than the originally proposed cubic $LuH_3$[2]. Pressure-induced color changes from blue through pink to bright red were indeed verified in both undoped[3] and N-doped $LuH_2$[4, 5], and the underlying mechanism has been revealed through the optical reflectivity measurements under high pressures[6]. However, for both undoped and N-doped $LuH_2$, no superconductivity was observed from 300 to 1.5 K under pressures up to 30 GPa[3-5], in striking contrast to the original report[1]. For the $LuH_{2\pm x}N_y$ samples prepared with the large-volume press, some kink- or hump-like features were observed in the temperature-dependent resistivity around room temperatures at high pressures[4, 5]. But these features are distinct and far weaker than those reported in Ref. [1] showing an abrupt drop in resistivity. It is noteworthy that caution should be given when interpretating the suspicious features in temperature-dependent resistance measured in a diamond anvil cell (DAC). Given the fact that the essential feature of resistivity drop in Ref. [1] has not been reproduced, the debate on the mystery of near-ambient superconductivity remains unsettled[4, 5].

Since the available experimental and theoretical investigations consistently ascertained that the actual chemical composition of the sample in Ref. [1] should be $LuH_2$ or N-doped $LuH_2$[2, 4, 5, 7], we decided to carry out comprehensive investigations on the parent compound $LuH_2$. During the course of our study, we noticed that the resistivity of cold-pressed (CP) $LuH_2$ pellets made of chemically purchased powders can vary sensitively upon modifying the grain size or surface conditions via the grinding process. Surprisingly, we can occasionally yet repeatedly observe abrupt resistivity drops at high temperatures, which also show dependences on magnetic fields and electrical current, in reminiscent of the observations in Ref. [1]. However, our detailed studies exclude the possible causes of structural, magnetic, or superconducting transition for the observed resistivity drop. Instead, we tentatively attribute it to the percolation of the metallic grains through the insulating layers on the grain surface, which are likely produced by the modification of hydrogen stoichiometry or the pollution by oxygen/nitrogen in air.

## Results

Because the hydrogenation process is usually employed to convert ductile rare-earth metals to brittle hydrides in order to facilitate the pulverization process, the main phase of most commercially purchased "rare-earth" powders are actually rare-earth dihydrides. In the present study, we used the commercially available "Lu" powder (99.9%, JiangXi Viilaa Metal Material Co., Ltd.) as the starting material. Rietveld refinements on the powder XRD pattern confirmed that the major phase is the cubic $LuH_2$ (95%) coexisting with minor Lu (5%) and some unidentified phases, as shown in Fig. S1. Thus, the as-received "Lu" powder is dark blue in color and contains mainly the $LuH_2$ grains with different sizes ranging from ~200 μm to submicron.

To evaluate the intrinsic transport properties of $LuH_2$, we picked up a large grain of ~200 μm in length from the as-received powder and measured its temperature-dependent resistivity, $\rho(T)$, by using the standard four-probe method. The inset of Fig. 1(a) shows the photograph of the measured grain attached with four gold leads by silver paste. As shown in Fig. 1(a), the $LuH_2$ grain displays a typical metallic behavior with a small room-temperature resistivity of ~9.3 ×$10^{-8}$ Ω m, which is on the same order of high-purity cooper. The resistivity decreases upon cooling down and saturates below ~20 K to a constant value of ~0.15 ×$10^{-8}$ Ω m, giving rise to a relatively large residual resistivity ratio RRR ≡ $\rho$(300 K)/$\rho$(2 K) = 62.7. We find that the $\rho(T)$ below 40 K can be described by the power law, viz. $\rho(T) = \rho_0 + AT^n$ with n = 3.5, as shown by the dotted line in Fig. 1(a). Both the absolute resistivity value and the RRR of the $LuH_2$ grain are comparable to those of the stoichiometric bulk sample reported previously[8, 9]. Our results thus confirmed that the blue-colored $LuH_2$ is characterized by an intrinsic metallic behavior with high conductivity comparable to simple metals.

When the as-received powder was directly cold pressed at 4 GPa into dense pellets by using a large-volume press, we find that its resistivity is enhanced by almost four orders of magnitude in comparison with that of the above $LuH_2$ grain. The samples made in this way are labeled as "As-received+CP" hereafter. We measured $\rho(T)$ of over ten samples and shown in Fig. 1(b) the $\rho(T)$ curves for four representative samples. Although the $\rho(T)$ still shows metallic behavior, the room-temperature resistivity has been increased significantly to the level of 2-5 × $10^{-4}$ Ω m, and the RRR value is reduced significantly to 1.1-1.3. As mentioned above, the as-received powder contains $LuH_2$ grains of different sizes, such an enhanced resistivity in the "As-received+CP" pellets can be attributed to the presence of submicron-sized grains, which contributed significantly to resistivity due to the grain boundary scattering. However, the enhancement of $\rho(T)$ by nearly four orders of magnitude is surprisingly large, which indicates that the surface of the $LuH_2$ grains scatters electrons strongly.

During the $\rho(T)$ measurements of "As-received+CP" LuH$_2$ samples, we occasionally observed in one sample a pronounced resistivity drop at about 200-250 K as shown in Fig. 2(a). In comparison with the typical samples shown in Fig. 1(b), this sample is featured by a larger resistivity of ~1.8 × 10$^{-2}$ Ω m at room temperature. For the first measurement upon cooling down under 0 T, its $\rho(T)$ starts to decrease quickly from ~250 K and then levels off below ~ 212 K, showing about a 10% drop in resistivity. This anomaly shows up repeatedly during the thermal cycling between 300 and 2 K, and it can even be shifted down by magnetic fields. At first glance, this anomaly seems to signal the occurrence of superconductivity. However, our following studies are against such a scenario. For this specific sample, this anomaly is rather robust and always shows up for many runs at different currents, Fig. 2(b). It is noticeable that the anomaly moves to higher temperature at larger current and the magnitude of resistivity also varies for different runs.

To verify if the observed resistivity drops corresponding to a superconducting transition, we measured the dc magnetic susceptibility of the exact same sample used for the above resistivity measurements. Before the susceptibility measurements, the electrical leads were detached, and the sample surface was slightly polished to remove the silver past. Unfortunately, we did not observe any feature due to the Meissner effect within the limit of our high-resolution SQUID. Then, we reattached electrical leads on the same sample to measure its resistivity again. To our surprise, the anomalous drop in resistivity disappears and cannot be reproduced on the identical sample. This puzzled us and motivated us to investigate the underlying mechanism with an aim to reproduce the resistance drops in the cold-pressed LuH$_2$.

As mentioned above, we noticed that the "As-received+CP" LuH$_2$ samples show a ten-thousand times larger resistivity and the specific sample showing resistivity drop possesses an even larger room-temperature resistivity in comparison with that of the bulk grain. We realized that it could be an effective way to further enhance the resistivity of LuH$_2$ by reducing the grain size or increasing the surface area. To test this hypothesis, we manually ground the as-received powder for 5 minutes in air with a mortar and pestle, and then cold-pressed it into dense pellets at 4 GPa as done above. This sample is labeled as "Ground+CP" to distinguish it from the above "As-received+CP" sample. As shown in Fig. 1(c), the $\rho(T)$ of most "Ground+CP" samples increase slightly upon cooling down, displaying a weakly localized non-metallic behavior. In addition, the room-temperature resistivity also increases further by one order of magnitude to ~7 × 10$^{-3}$ Ω m. This result demonstrated that the additional grinding process is effective in modifying the transport properties of LuH$_2$ samples.

Interestingly, for some "Ground+CP" LuH$_2$ samples, we can reproducibly observe the sharp resistivity drop as shown in Fig. 3 for a representative example. The panels in Fig. 3 are displayed in the same order as the measurement sequences. As seen in Fig. 3(a), the $\rho(T)$ at 0 T exhibits an abrupt drop at 50 K during the first cooling-down process while it jumps back at ~ 200 K during the warming-up process, resulting in a significant thermal hysteresis. Upon increasing magnetic field to 7 T at room temperature and measuring $\rho(T)$ again, Fig. 3(b), the resistivity anomaly takes place at ~160 K and 210 K for the cooling-down and warming-up processes, respectively. The positive field effect thus excludes the possibility of superconductivity. When the field is reduced to zero and the current is increased from 20 to 100 μA, Fig. 3(c), the anomalous drop in resistivity still stays at high temperatures of 150 and 230 K with a slightly wider hysteresis. Finally, when the current is reduced to the original 20 μA, Fig. 3(d), however, the hysteresis becomes much reduced while keeping the resistivity drop at 150-170 K. Similar behaviors can be reproduced on another "Ground+CP" sample, Fig. S2, but the temperatures of the resistivity anomalies are different. In comparison with the "As-received+CP" samples, the "Ground+CP" samples with larger resistance show a high probability to present resistivity drops. However, such an abrupt drop anomaly in resistivity does not follow any regular dependences on magnetic fields or electrical currents according to our results.

To clarify the origin of such a resistivity-drop anomaly and its strong dependence on the process conditions, we first checked the powder XRD at room temperature for these three samples. As shown in Fig. S3, both the XRD patterns and the calculated lattice parameters for these samples are almost identical, while the peaks of the CP samples are slightly broadened up due to the presence of stress/strain introduced by the grinding and cold-pressing processes. Then, we further measured variable-temperature XRD of the ground powder from 300 down to 100 K, Fig. 4(a). The absence of any peak splitting and smooth evolution of lattice parameter as a function of temperature rule out the possibility of structural phase transition for $LuH_2$ in the investigated pressure range. Linear fitting to the temperature dependence of unit-cell volume $V(T)$ yields a thermal expansion coefficient (TEC) of $\alpha = 11.98 \times 10^{-6}$ K$^{-1}$, or a linear TEC of $\alpha_L = 3.98 \times 10^{-6}$ K$^{-1}$, Fig. 4(b). We also measured the temperature dependence of magnetic susceptibility for these samples, Fig. S4, which show similar paramagnetic behaviors without discernable anomalies in the temperature range 2-350 K, thus ruling out the possibility of magnetic transition as the cause of the sudden resistivity drop.

Then, we probe the intrinsic electronic properties of the CP samples made from as-received and ground powders via specific-heat measurements. As seen in Fig. 5(a), except for a dip anomaly around 220 K associated with the Apiezon N grease, no obvious anomalies can be discerned on the $C(T)$ curves in the whole temperature range, further confirming the absence of structural, magnetic, or superconducting transitions. It is noted that the $C(T)$ of the "Ground+CP" sample is slightly larger than that of the "As-received+CP" sample, especially at high temperatures. The $C(T)$ data below 7 K was replotted in Fig. 5(b) in the form of $C/T$ vs $T^2$, which follows nicely a linear behavior due to the lattice and electronic contributions. A linear fitting to the $C/T$ vs $T^2$ data yields almost identical Sommerfeld coefficient $\gamma = 2.1$ mJ mol$^{-1}$ K$^{-2}$ for these two samples, even though their $\rho(T)$ data in Fig. 1(b, c) show distinct behaviors. Such a small $\gamma$ is consistent with the intrinsic metallic ground state for the bulk LuH$_2$ samples, while the observed weak localization of $\rho(T)$ for the "Ground+CP" sample should be mainly attributed to the strong scattering of electrons by the insulating layers on the grain surfaces.

Although the LuH$_2$ is stable at ambient conditions, the surfaces of LuH$_2$ can be easily damaged or contaminated by oxygen/nitrogen, resulting in an insulating layer on the grain surfaces according to a previous study[9]. Thus, operations of Lu and LuH$_2$ were usually carried out under mineral oil[10]. In the present study, however, all operations were performed in air, which can inevitably produce insulating layers due to the modification of hydrogen stoichiometry and/or pollution by oxygen/nitrogen on the grain surfaces. This effect is expected to be magnified and expeditated through the additional grinding process, in consistent with the observations of much enhanced resistivity in the "Ground+CP" sample shown in Fig. 1(c). To elaborate this point, we elongated the grinding time to 30 min and indeed observed further enhancement of resistivity in comparison with that ground for 5 min, as shown in Fig. S5. The incorporation of heavier oxygen/nitrogen on the surfaces can also partially account for the smaller Debye temperature for the "Ground+CP" sample with respect to that of the "As-received+CP" sample, shown in Fig. 5(b).

Therefore, the present work demonstrates that the CP LuH$_2$ samples consist of the metallic grains with high conductivity and the insulating surfaces, and their relative ratio can be modified through the grinding process in air. For the "As-received+CP" sample, the metallic grains still dominate the transport properties and retain the metallic behavior in $\rho(T)$ as observed in Fig. 1(b). When the contributions from the insulating surfaces are enhanced by grinding the as-received powder, the "Ground+CP" samples usually display a weakly localized $\rho(T)$ behavior as seen in Fig. 1(c). Under the right conditions, percolations of the metallic grains through the insulating surfaces can give rise to sudden drops of resistivity upon varying temperatures. As a probability issue, the percolation takes place randomly and it can thus rationalize the observed irregularities of the resistivity drop in the studied "Ground+CP" sample. It seems that the CP LuH$_2$ sample made of as-received powder after grinding for 5 minutes gives a high rate to reproduce the resistivity drop for our studied samples. For those CP samples made of well-ground powders, the contributions from the insulating layers become dominant, preventing the occurrence of percolation.

Based on the above studies, we can conclude that the observed resistance drops in the CP LuH$_2$ samples has nothing to do with superconductivity, magnetic, or structural phase transitions, but is mostly likely attributed to the percolation of metallic grains through the insulating surfaces. We speculate that the insulating surfaces of LuH$_2$ originate from the modification of hydrogen stoichiometry or the contamination by oxygen/nitrogen during the grinding process in air. The reduced Debye temperature on the "Ground+CP" sample is consistent with this scenario. More studies are needed to clarify the mechanism of surface degradation in $LuH_2$. If this scenario is correct, intermixing of highly conducting $LuH_2$ with other insulating phases such as LuN can also produce sudden drops in resistivity. The present study thus demonstrates that an abrupt drop in resistivity does not mean the occurrence of superconductivity, which call for caution in interpreting the anomalies in resistivity. In this case, it is definitely invalid to perform background subtraction in treating the resistivity data.

## Conclusion

In summary, we demonstrated that the resistivity of polycrystalline $LuH_2$ is very sensitive to the grain conditions and can be modified from metallic to weakly localized behavior through a simple grinding process in air. By controlling the grinding time, we can repeatedly observe the abrupt resistivity drop at high temperatures in the "Ground+CP" $LuH_2$ samples. We proposed that it is the percolation of the metallic grains through the insulating layers on the grain surfaces that produces the resistivity drops. Our present results call for caution in asserting the resistivity drop as superconductivity and invalidate the background subtraction in analyzing the resistivity data.

## Acknowledgments

This work is supported by the National Natural Science Foundation of China (Grant Nos. 12025408, 11921004, 11888101), the Beijing Natural Science Foundation (Z190008), the National Key R&D Program of China (2021YFA1400200), the Strategic Priority Research Program of CAS (XDB33000000).

## References

[1] Dasenbrock-Gammon N, Snider E, McBride R, Pasan H, Durkee D, Khalvashi-Sutter N, Munasinghe S, Dissanayake S E, Lawler K V, Salamat A and Dias R P 2023 *Nature* **615** 244

[2] Liu M, Liu X, Li J, Liu J, Sun Y, Chen X-Q and Liu P 2023 *ArXiv:2303.06554*

[3] Shan P, Wang N, Zheng X, Qiu Q, Peng Y and Cheng J 2023 *Chin. Phys. Lett.* **40** 046101

[4] Zhang Y-J, Ming X, Li Q, Zhu X, Zheng B, Liu Y, He C, Yang H and Wen H 2023 *Sci. China-Phys. Mech. Astron.* **66** 287411

[5] Xing X, Wang C, Yu L, Xu J, Zhang C, Zhang M, Huang S, Zhang X, Yang B, Chen X, Zhang Y, Guo J-g, Shi Z, Ma Y, Chen C and Liu X 2023 *ArXiv2303.17587*

[6] Zhao X, Shan P, Wang N, Li Y, Xu Y and Cheng J 2023 *ArXiv:2303.16745*

[7] Xie F, Lu T, Yu Z, Wang Y, Wang X, Meng S and Liu M 2023 *ArXiv:2303.11683*

[8]  Daou J N, Lucasson A, Vajda P and Burger J P 1984 *J. Phys. F: Met. Phys.* **14** 2983
[9]  Daou J N, Vajda P, Burger J P and Shaltiel D 1988 *EPL* **6** 647
[10] Pebler A and Wallace W E 1962 *J. Phys. Chem.* **66** 148

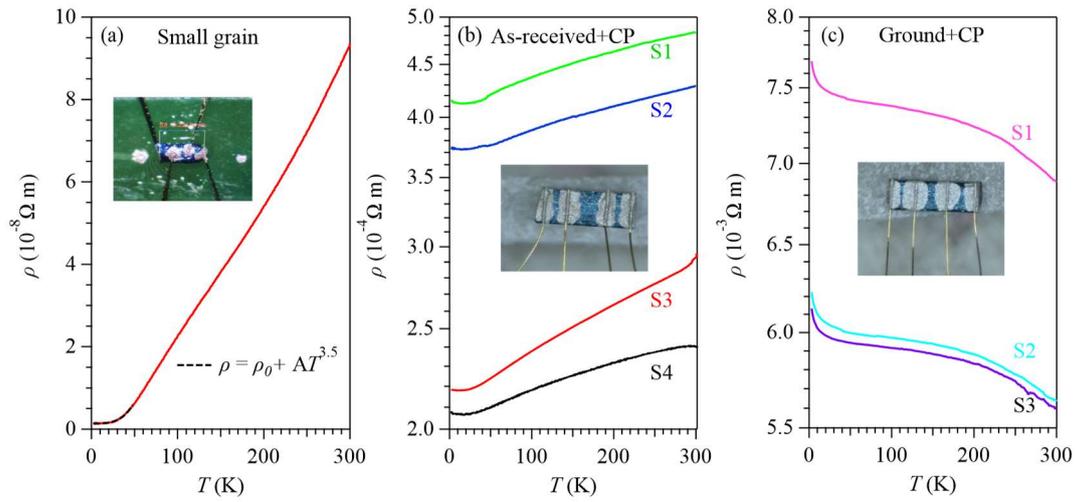

**Figure 1**. Temperature-dependent resistivity $\rho(T)$ of several $LuH_2$ samples: (a) the small grain picked up from the as-received powder; (b) the cold-pressed samples made of the as-received powder; (c) the cold-pressed samples from the ground $LuH_2$ powder. The inset shows the samples' photographs for measuring resistivity.

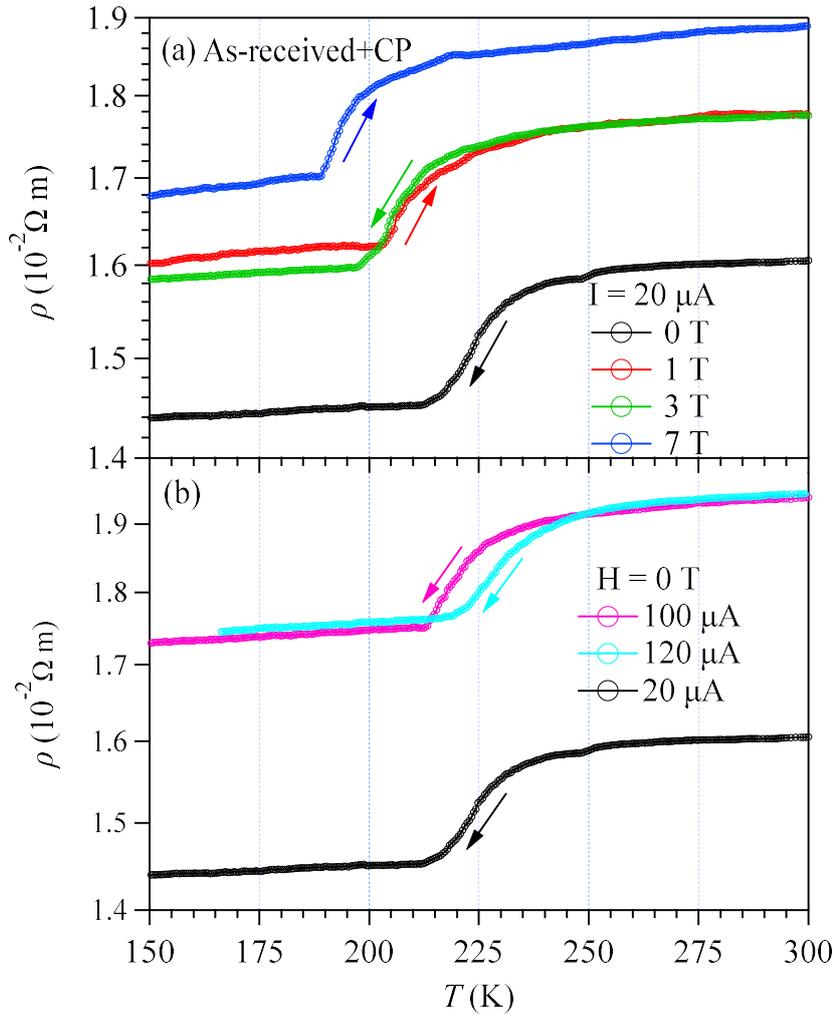

**Figure 2**. Temperature-dependent resistivity $\rho(T)$ of one "As-received+CP" LuH$_2$ sample that shows resistivity drop: (a) measured at different magnetic fields with the same electrical current, and (b) measured at zero field with different electrical currents.

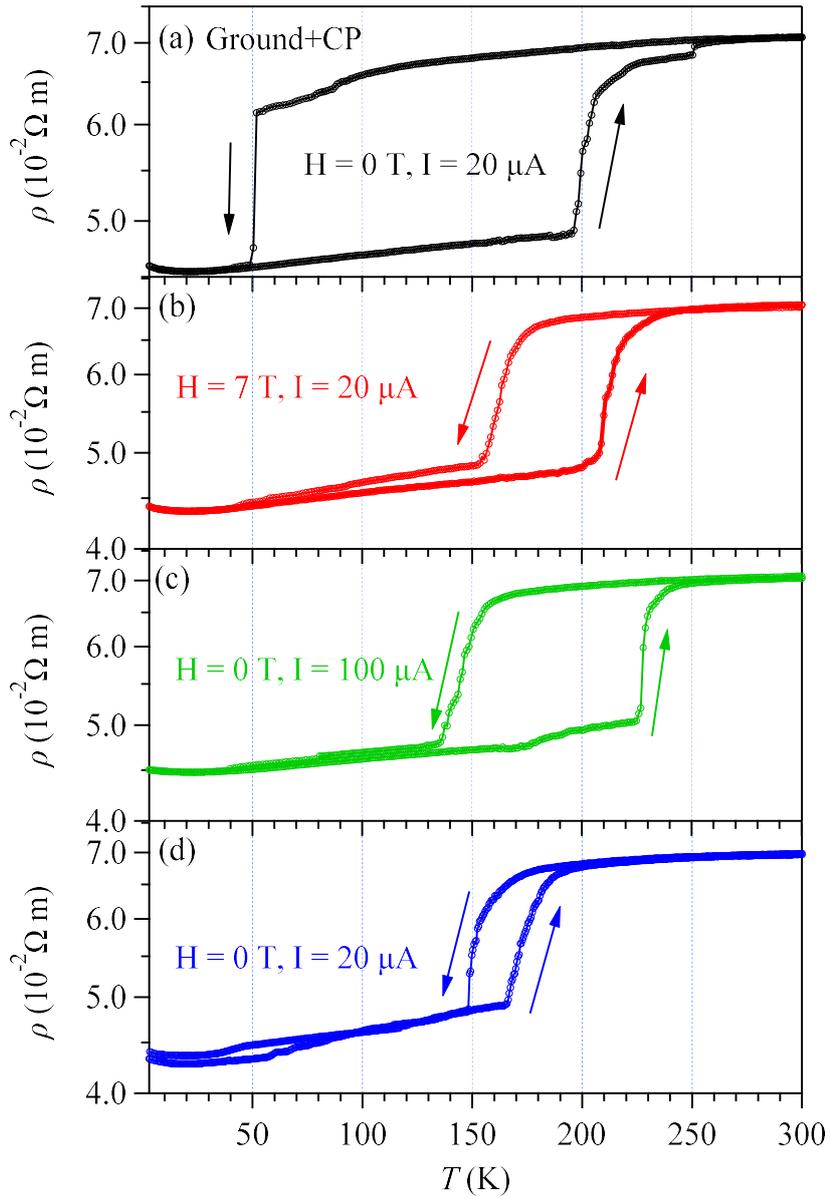

**Figure 3** Temperature-dependent resistivity $\rho(T)$ of one "Ground+CP" LuH$_2$ sample with 5-min grinding. The panels are organized in the same order as the measurement sequences. The arrows indicate the cooling-down and warming-up processes.

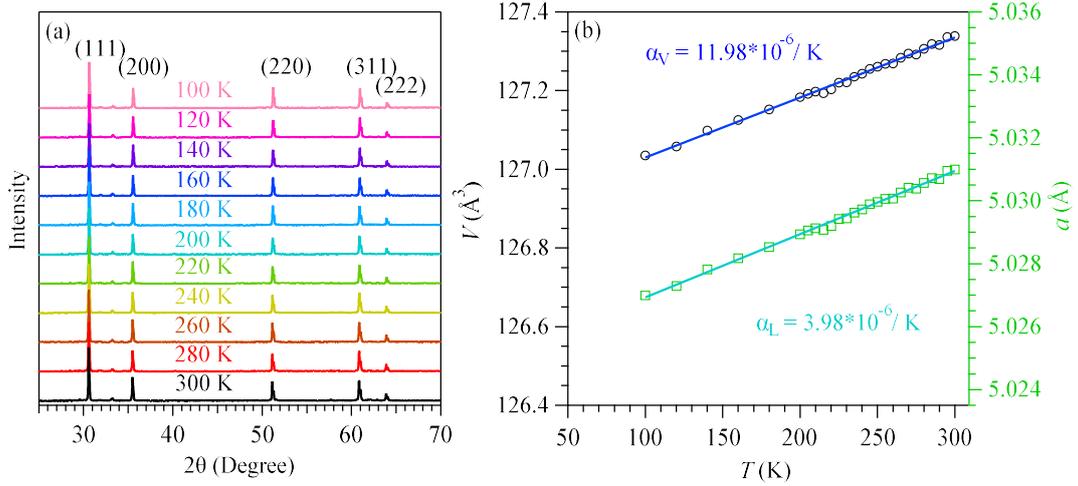

**Figure 4.** (a) Variable-temperature XRD of the ground powder from 300 down to 100 K. (b) Temperature dependences of unit-cell volume and lattice parameter with linear fitting curves shown by solid lines.

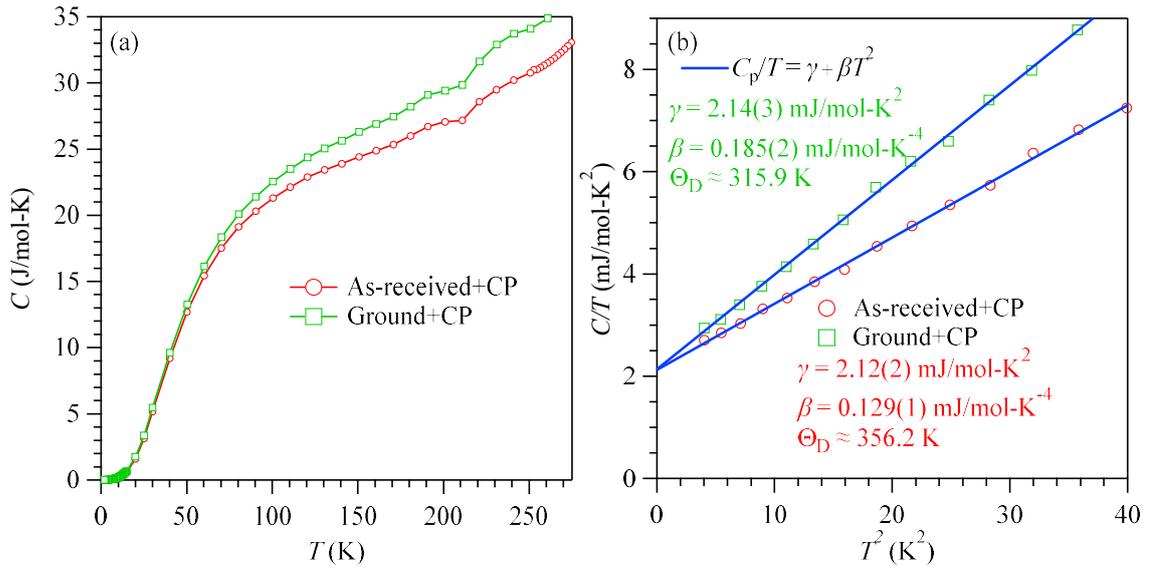

**Figure 5** (a) Specific heat $C(T)$ for two CP LuH$_2$ samples without and with additional grinding in the wide temperature range from 2 to 270 K under zero field. (b) The plot of $C/T$ vs $T^2$ in the low-temperature range. A linear fitting was applied to extract the Sommerfeld coefficient $\gamma$ and the Deby temperature $\Theta_D$.

# Supplementary Materials

# Percolation-induced resistivity drop in cold-pressed LuH$_2$

Ningning Wang[1,2#*], Jun Hou[1,2#], Ziyi Liu[1,2], Pengfei Shan[1,2], Congcong Chai[1,2], Shifeng Jin[1,2], Xiao Wang[1,2], Youwen Long[1,2], Yue Liu[1,2], Hua Zhang[1,2], Xiaoli Dong[1,2], and Jinguang Cheng[1,2*]

[1]Beijing National Laboratory for Condensed Matter Physics and Institute of Physics, Chinese Academy of Sciences, Beijing 100190, China

[2]School of Physical Sciences, University of Chinese Academy of Sciences, Beijing 100190, China

\# These authors contributed equally to this work.

*Corresponding authors: nnwang@iphy.ac.cn; jgcheng@iphy.ac.cn

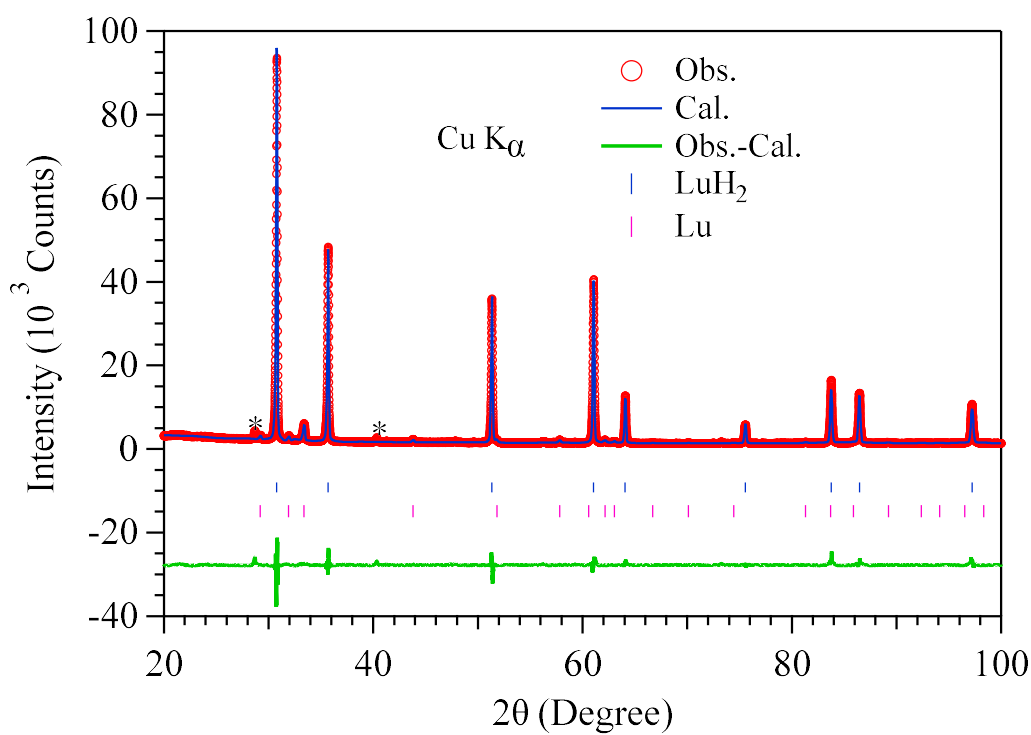

**Figure S1.** Observed (circle), calculated (solid line), and difference (bottom line) XRD profiles of the commercially purchased "Lu" powder measured at room temperature after Rietveld refinements. The Bragg positions are shown as the tick marks. Two additional weak peaks at 28.70° and 40.35° marked by asterisks come from some unknown phases, which thus cannot be considered in the Rietveld refinement.

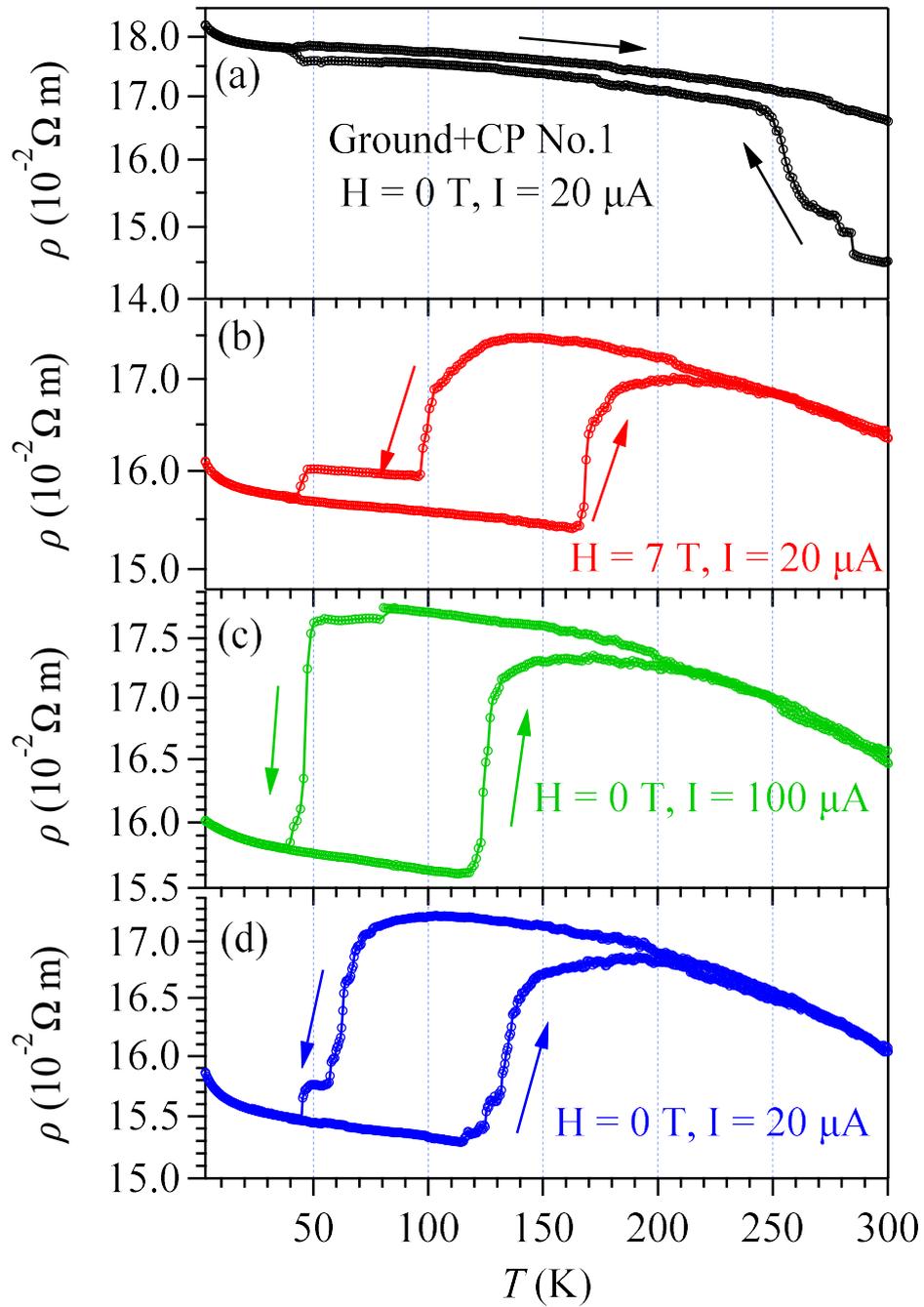

**Figure S2.** Temperature-dependent resistivity $\rho(T)$ of another "Ground+CP" $LuH_2$ showing a drop/jump behavior. The arrows indicate the cooling-down and warming-up processes.

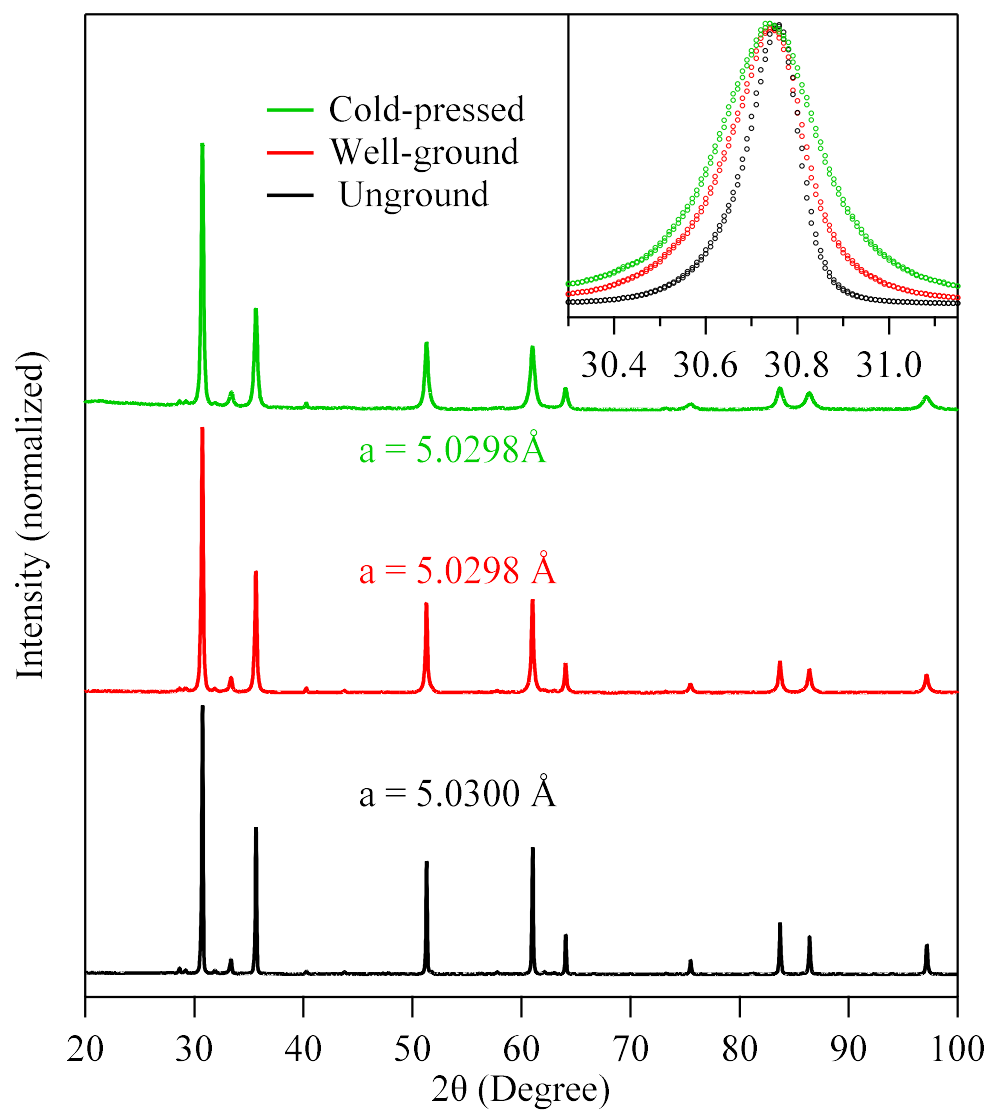

**Figure S3**. Powder XRD patterns of three kinds of LuH$_2$ samples with different treatments: as-received powder (bottom), ground powder (middle), and powder of CP sample (top). The inset shows the main peak of these LuH$_2$ samples.

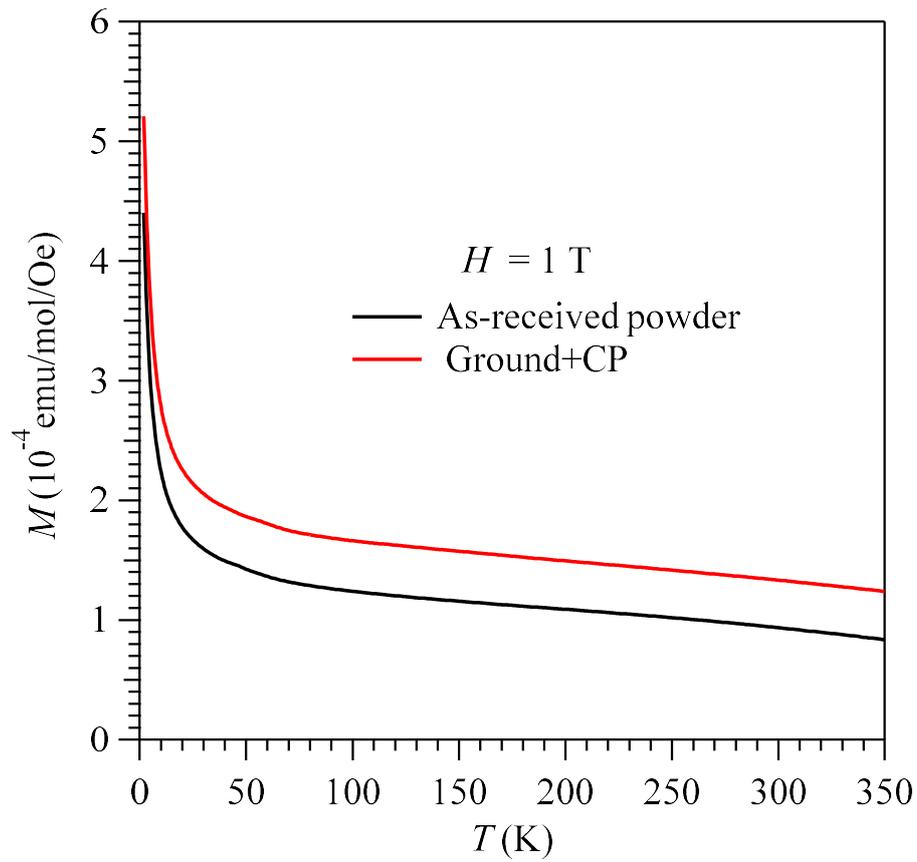

**Figure S4.** Temperature-dependent dc magnetic susceptibility $\chi(T)$ of the as-received powder and the "Ground+CP" samples of LuH$_2$.

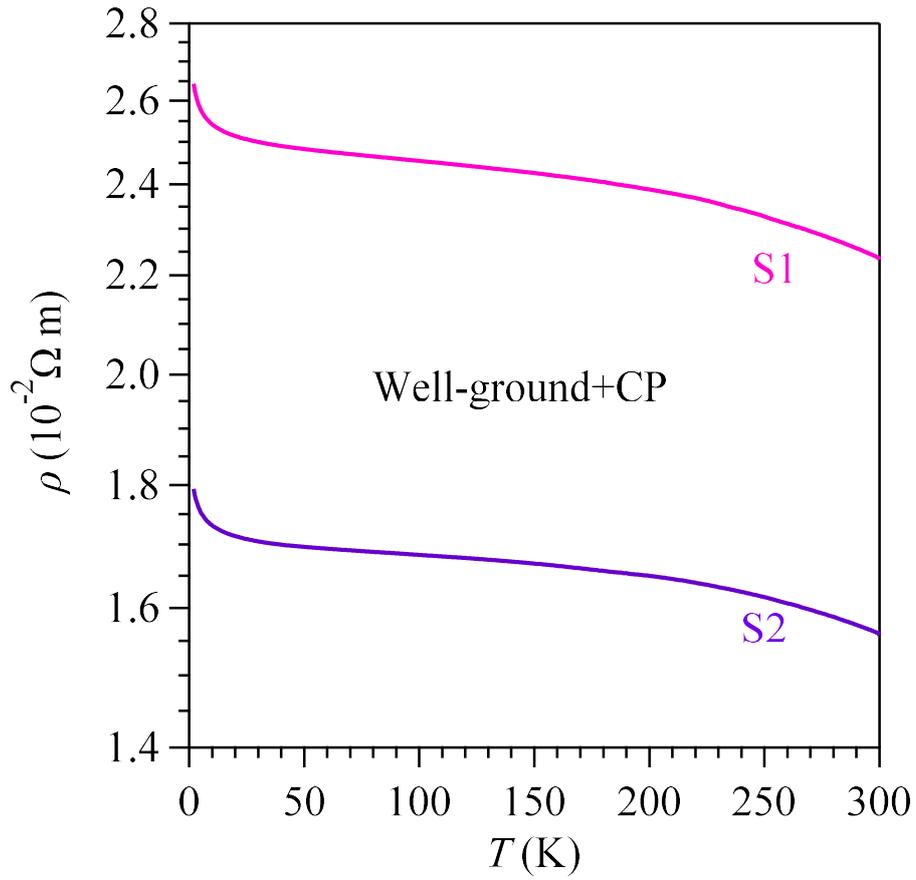

**Figure S5.** Temperature-dependent resistivity $\rho(T)$ of the CP LuH$_2$ sample made of as-received powders after grinding for 30 minutes.